\pdfoutput=1
\documentclass[12pt,epsf]{article}
\usepackage{graphicx}
\usepackage{cite}
\usepackage{amsmath} 
\usepackage{color}

\newcommand{\beq}{\begin{equation}}
\newcommand{\eeq}{\end{equation}}
\newcommand{\bea}{\begin{eqnarray}}
\newcommand{\eea}{\end{eqnarray}}

\newcommand{\MP}{\overline{m}_\mathrm{P}}

\newcommand{\Treh}{T_\mathrm{reh}}

\newcommand{\CP}{\mathcal{CP}}
\newcommand{\C}{\mathcal{C}}
\newcommand{\B}{\mathcal{B}}
\begin{document}
\thispagestyle{empty}
\vspace*{-15mm}
{\bf OCHA-PP-342}\\
\vspace{10mm}
\begin{center}
{\Large\bf
Ratchet Baryogenesis with an Analogy to the Forced Pendulum \\
}
\vspace{7mm}

\baselineskip 18pt
{\bf Kazuharu Bamba${}^{1*}$,
Neil D. Barrie${}^{2\dagger}$,
Akio Sugamoto${}^{3, 4\ddagger}$,\\
Tatsu Takeuchi${}^{5\S}$, 
Kimiko Yamashita${}^{3, 6\P}$
}
\vspace{1mm}

{\it
${}^{1}$Division of Human Support System, Faculty of Symbiotic Systems Science,\\
Fukushima University, Fukushima 960-1296, Japan\\
${}^{2}$ARC Centre of Excellence for Particle Physics at the Terascale,\\
School of Physics, The University of Sydney, NSW 2006, Australia\\
${}^{3}$Department of Physics, Graduate School of Humanities and Sciences,\\
Ochanomizu University, 2-1-1 Ohtsuka, Bunkyo-ku, Tokyo 112-8610, Japan\\
${}^{4}$Tokyo Bunkyo SC, the Open Universtiy of Japan, Tokyo 112-0012, Japan \\
${}^{5}$Center for Neutrino Physics, Department of Physics,\\Virginia Tech,
Blacksburg VA 24061, USA\\
${}^{6}$Program for Leading Graduate Schools, \\
Ochanomizu University, 2-1-1 Ohtsuka, Bunkyo-ku, Tokyo 112-8610, Japan\\
${}^{*}$bamba@sss.fukushima-u.ac.jp,
${}^{\dagger}$neil.barrie@sydney.edu.au,
${}^{\ddagger}$sugamoto.akio@ocha.ac.jp,
${}^{\S}$takeuchi@vt.edu,
${}^{\P}$yamashita@hep.phys.ocha.ac.jp}
\end{center}
\begin{center}
\begin{minipage}{14cm}
\baselineskip 16pt
\noindent
\begin{abstract}%
A new scenario of baryogenesis via the ratchet mechanism is proposed based on an analogy to the forced pendulum.
The oscillation of the inflaton field 
during the reheating epoch after inflation plays the role of the driving force, 
while the phase $\theta$ of a scalar baryon field (a complex scalar field with baryon number) plays the role of the angle of the pendulum. 
When the inflaton is coupled to the scalar baryon, the behavior of the phase $\theta$ can be analogous to that of the angle of the forced pendulum.
If the oscillation of the driving force is adjusted to the pendulum's motion, a directed rotation of the pendulum is obtained with a non-vanishing value of $\dot{\theta}$, which models successful baryogenesis since $\dot{\theta}$ is proportional to the baryon number density.
Similar ratchet models which lead to directed motion have been used in the study of molecular motors in biology. 
There, the driving force is supplied by chemical reactions, while in our scenario this role is played by the inflaton during the reheating epoch.
\end{abstract}
\end{minipage}
\end{center}

\baselineskip 18pt
\def\thefootnote{\fnsymbol{footnote}}
\setcounter{footnote}{0}



\newpage

The objective of baryogenesis is to explain the small but non-zero baryon asymmetry of the universe, which according to the 2015 Planck
analysis \cite{Ade:2015xua} stands at 
$\eta_{B}=n_B/s \simeq~8.5 \times 10^{-11}$,
where $n_B$ and $s$ are respectively the baryon number and entropy densities of the universe.
The Standard Model (SM) of particle physics is unable to explain this value, due to the $\CP$ violation present within the SM being too small,\footnote{%
$\CP$ asymmetry in the SM is suppressed by the Jarlskog factor coming from the mass differences of quarks, $\prod_{i < j}(m_{u_i}^2-m_{u_j}^2)(m_{d_i}^2-m_{d_j}^2)/(100~\mbox{GeV})^{12} \approx 10^{-13}$, where $(i, j)$ stand for the generation indices.}  
and thus some new source of $\CP$ violation is required \cite{Cohen:1993nk}.
It is usually assumed that the baryon asymmetry at the end of the inflationary epoch was negligibly small or zero, 
due to the rapid dilution of any initial baryon number density that may have existed. 
Due to this, most mechanisms of baryogenesis are assumed to occur after inflation; during the reheating or subsequent epochs prior to Big Bang Nucleosynthesis (BBN).

In this letter, we propose a baryogenesis-during-reheating  model which takes its inspiration from 
the ``ratchet'' models of molecular motors in biological systems \cite{Reimann:1996,Takeuchi:2010tm},
e.g. the motion of myosin molecules along actin filaments.
The question in biological motor theory is how the motors achieve directed motion, 
e.g. the myosin molecule ``walks'' from one end of the actin filament to the other, instead of random walking back and forth.  
This type of directed motion can be modeled by assuming that  
the molecular motor, the position of which we denote by $\theta$, moves within a periodic potential $V(\theta)$
which is asymmetric under the reflection $\theta \to -\theta$, \textit{i.e.} a ``ratchet'' potential.
The force which drives the molecular motor is supplied quasi-periodically by chemical reactions such as ATP $\to$ ADP.
Alternatively, one could also model directed motion with a reflection-symmetric potential, but assuming that the interaction with the driving force
breaks the symmetry, a type of ``temporal ratchet.''
The baryogenesis model we propose here utilizes the latter type of ratchet mechanism, 
in which the driving force is supplied by the oscillation of the inflaton,
the position of the motor is embodied in the phase $\theta$ of a complex scalar field that carries baryon number,
and the required breaking of the reflection symmetry (which corresponds to $\CP$ violation) is realized via the coupling of the
inflaton to the scalar baryon.

Thus, our model
is a coupled system of the inflaton field and a complex scalar field 
in an expanding universe with scale factor $a(t)$.  

Models of interest to our work have been proposed in the past. The model considered by Dimopoulos and Susskind in Ref. \cite{Dimopoulos:1978kv} could be seen as a prototype of the ratchet baryogenesis model, given that they utilised a complex scalar field carrying a baryon number.  The model, however, gives a non-zero baryon number ($\dot{\theta}\ne 0$) only when the scalar potential is non-renormalizable, that is, in case of $V(\phi)=\lambda (\phi^{*}\phi)^n(\phi+\phi^{*})(\alpha \phi^3+\alpha^*\phi^{*3})$ with $n > 1/4$.  An advantage of our mechanism is that we consider a renormalizable one~\cite{Takeuchi:2010tm}. 

One of the most popular models within this framework is that of Affleck and Dine~\cite{Affleck:1984fy} in which the baryon number is effectively produced in Supersymmetric Grand Unified Theories (SUSY GUTs).  SUSY is assumed to be unbroken in the early universe, and the associated scalars $\phi$ (squarks and sleptons) can take large real or imaginary values due to the existence of a number of flat directions in the scalar potential.  After SUSY is broken at a lower energy, the scalars have mass $m$, and the fields start to roll down their potentials $V(\phi)=\frac{1}{2}m^2 \phi^2$.  If the real part $\Re(\phi)$ and the imaginary part $\Im(\phi)$ evolve differently, then a net baryon number is produced.  Affleck-Dine baryogensis is similar to the case we consider here, since both scenarios use the temporal development of scalar fields that carry baryon number and/or lepton number, but there are some differences.  First, our scenario does not depend on SUSY GUTs, and so is applicable to various models.  Secondly, in our scenario the external motive force is supplied by the rolling down and the oscillation by the inflaton field.  This role is carried by a specific component of the scalar fields in~\cite{Affleck:1984fy}.  Thirdly, in our model, since the external motive force (the pushing force of the pendulum from outside) is separated from the scalar baryon, we can find the ``phase locked state'', a known phenomena in the forced pendulum or Josephson current, which is a new ingredient of baryogenesis proposed here, and the mechanism for $\dot{\theta} \ne 0$ is also clarified.

Dolgov and Freese~\cite{Dolgov:1994zq} also introduced a complex scalar field with a baryon number, and studied the temporal development of its phase $\theta$ which was identified as a Nambu-Goldstone mode.  So, the mechanism is similar to ours, and the scalar field they consider is analogous to our scalar baryon field $\phi(t)$. In their case the dynamics do not, however, go beyond the linear approximation of the damped harmonic oscillator, although the mass and the damping factor are, respectively, modified by and given by the back reaction of fermions. In our model the external pushing force is separately introduced, and hence the ``phase locked state'' is found to play an important role in the baryogenesis. 


Now, we will come back to our model in which the inflaton is denoted by a real scalar field $\Phi(t)$ with a potential $U(\Phi)$, 
while the scalar baryon is denoted by $\phi(t)=\frac{1}{\sqrt{2}} \phi_r e^{i \theta(t)}$ with a potential $V(\phi)=\lambda\phi^*\phi(\phi-\phi^*)(\phi^*-\phi)=\lambda\phi^4_r\sin^2\theta(t)$, where $\phi_r$ is assumed to be constant. 
%
If the charge of the complex field is identified as baryon number, 
then the baryon number symmetry $\B$ becomes the rotational symmetry in the angle $\theta$,  
$\B: \theta(t) \to \theta(t) + \mathrm{const}$, and it is violated by this potential.  
The corresponding baryon number current is 
$j_{B}^{\mu}=i( \phi \partial^{\mu}\phi^{*} -\phi^{*}\partial^{\mu}\phi)=\phi_{r}^2 \partial^{\mu} \theta$, and the baryon number density is $n_B=\phi_{r}^2 \dot{\theta}$.
The charge conjugation symmetry $\C$ is given by $\C: \phi \to \phi^*$, or $\theta \to -\theta$, so $\C$ is conserved in this potential.  
This point differs from the usual ratchet models that have reflection-asymmetric potentials.

Here, the spatial dependence of the scalar fields is ignored, since we consider a uniform and isotropic universe.  
This means that the parity symmetry $\mathcal{P}$ is always conserved.  
So the potential $V(\phi)$ violates $\B$, but preserves $\C$ and $\CP$.  
The following coupling between the inflaton and the scalar baryon is introduced: $\mathcal{L}_{int}=-\frac{1}{\Lambda} j_B^{\mu} \partial_{\mu} \Phi=-\frac{\phi_r^2}{\Lambda} \partial^{\mu} \theta \partial_{\mu} \Phi$, which is a natural way to introduce interactions when current conservation is broken.  This type of interaction appears in the coupling of nucleons to pions when chiral symmetry is broken, 
and has also been used by Cohen and Kaplan in their model for baryogenesis~\cite{Cohen:1987vi}.  
This interaction has dimension five, so the mass scale $\Lambda$ must be introduced, and it violates $\C$ and $\CP$.

Now, the action of our model reads
\begin{eqnarray}
S = \int dt \, a(t)^3\,
\biggl[\,
\frac{\phi_r^2}{2}\, \dot{\theta}^2
\,-\,\lambda\phi_r^4\sin^2\theta
+\, \frac{1}{2}\,\dot{\Phi}^2
\,-\,U(\Phi) \,-\, 
\frac{\phi_r^2}{\Lambda}\,\dot{\theta} \dot{\Phi} 
\,\biggr]
\;.
\label{eq:action2}
\end{eqnarray}
There are a number of possibilities in the choice of the inflaton potential $U(\Phi)$.  
We choose the Starobinsky potential~\cite{Starobinsky:1980te} due to its consistency with the observations of Planck \cite{Ade:2015xua}:
\begin{equation}
U(\Phi)
\;=\; \frac{3\mu^2\MP^2}{4}\left(1-e^{-\sqrt{2/3}\,\Phi/\MP}\right)^2
\;=\; \frac{1}{2}\mu^2\Phi^2 + \cdots
\;,
\label{eq:inf_potential}
\end{equation}
where $\MP=1/\sqrt{8\pi G_N}=2.4 \times 10^{18}$~GeV is the reduced Planck mass, and $\mu = (1.3 \times 10^{-5})\,\MP=3 \times 10^{13}$~GeV.

The Sakharov conditions must be satisfied for a successful baryogenesis mechanism. These criteria are the violation of $\B$, $\C$, and $\CP$, as well as a push-out from thermal equilibrium~\cite{Sakharov:1967dj}.  The model action we consider violates $\B$, $\C$, and $\CP$, so we now need to introduce into our model a push-out from thermal equilibrium. 
We will realize this during the reheating epoch after inflation, during which the inflaton oscillates in its potential. 
During this epoch the inflaton loses energy through Hubble damping and the decay into particles (this effect is embodied in the $\Gamma$ term below). 
The inflaton's energy is deposited into producing entropy (particles), 
and once this decay rate becomes faster than the Hubble expansion, the reheating epoch ends and the radiation dominated era begins.  
To account for this effect, we introduce a friction term to
the equations of motion derived from our action Eq.~(\ref{eq:action2}):
\begin{eqnarray}
\label{eq:diff_eq}
\ddot{\Phi} + \left(3H + \Gamma\,\right) \dot{\Phi} 
+ \frac{dU(\Phi)}{d\Phi} 
- \frac{\phi_r^2}{\Lambda}\bigl(\,\ddot\theta + 3H\dot\theta \,\bigr) 
& = & 0\;, \vphantom{\bigg|}
\cr
\ddot\theta + 3H\dot\theta + \lambda\phi^2_r\sin(2\theta)
-\frac{1}{\Lambda}\bigl(\,\ddot\Phi + 3H\dot\Phi \,\bigr)  
& = & 0\;. \vphantom{\bigg|}
\end{eqnarray}
The friction term $\Gamma\,\dot{\Phi}$ ensures that the necessary entropy is produced and the reheating epoch will come to an end,
but is assumed to not dominate the dynamics. 
The Hubble parameter $H = \dot{a}/a$ above takes on the time-dependence $\frac{2}{3t}$, the same as in the matter-dominated era.

We assume that $\phi_r^2/\Lambda \ll \langle \Phi \rangle$ holds. It follows that the inflaton motion is not affected by the scalar baryon, 
and its motion can be solved independently. 
That is, though the motion of the scalar baryon is driven by the ``external force" supplied by the inflaton, the recoil of the inflaton is
negligible small.
During the reheating epoch, the oscillation of the massive inflaton gives rise to a Hubble parameter which evolves as in the matter dominant era, and the inflaton undergoes damped oscillation in the approximately quadratic term $\frac{1}{2}\mu^2 \Phi^2$ in the Starobinsky potential, that is, 
\begin{eqnarray}
\ddot{\Phi} + \left(\frac{2}{t} + \Gamma \right) \dot{\Phi}+ \mu^2 \Phi=0,
\end{eqnarray}
which gives an analytic solution for $\Gamma \ll \mu$:
\begin{equation}
\Phi(t)
\;=\; \Phi_i \left(\frac{t_i}{t}\right)e^{-\Gamma(t-t_i)/2}\cos\left[\mu(t-t_i)\right]\;,
\label{PhiSolutionT}
\end{equation}
where $t_i$ is the beginning time of the reheating epoch, defined by the initial Hubble rate when the slow roll parameters are violated.

As for the motion of $\theta$, it can be described by
\begin{eqnarray}
\ddot{\theta}+f(t)\dot{\theta}
+ p(t) \sin(2\theta)
= - q(t)\cos\bigl[\mu(t-t_i)\bigr],
\label{thetaeom}
\end{eqnarray}
where
\begin{equation}
p(t) \;=\; \lambda\phi_r^2\;,~~f(t)=\frac{2}{t}, ~~
q(t) \;=\; \frac{\mu^2\Phi_i}{\Lambda}
\left(\frac{t_i}{t}\right)e^{-\Gamma(t-t_i)/2}\;.
\end{equation}
We now observe that the motion of $\theta$ is identical to that of a ``forced pendulum.'' 
The term proportional to $\sin(2\theta)$ can be viewed as the gravitational force on the pendulum, or swing, 
when it is at an angle $2\theta$ from the vertical.  
There is an added complexity, in that the strength of the external (pushing) force $q(t)$ and the friction $f(t)$ on the pendulum depend on $t$.


Baryogenesis is realized in this scenario when the solution of the equations of motion is found to give $\dot{\theta}(t_f) \ne 0$, at the end of reheating $t_f$. 
To obtain such a solution we consider the analogy to the forced pendulum, which tells us that we must adjust the timing and intensity of the external pushing to match the motion of the pendulum. 
In our scenario this corresponds to $p(t_f)\approx q(t_f)$, which imposes a condition on the input parameters to obtain driven motion at the end of reheating. We call this the ``sweet spot condition.''   
If the adjustment succeeds, the rotational motion of the pendulum around the fixed point arises with an almost constant angular velocity $\dot{\theta}$.  This is a solution which gives a non-zero baryon number in the universe, in which the inflaton plays a role of the pusher (or the parent), and the phase of the scalar baryon plays the role of the angle of the pendulum or the swing (on which the child is sitting).  
Therefore, this is a baryogenesis scenario based on an analogy to the forced pendulum.

\begin{figure}[ht]
\center
 \includegraphics[width=.495\textwidth,clip]{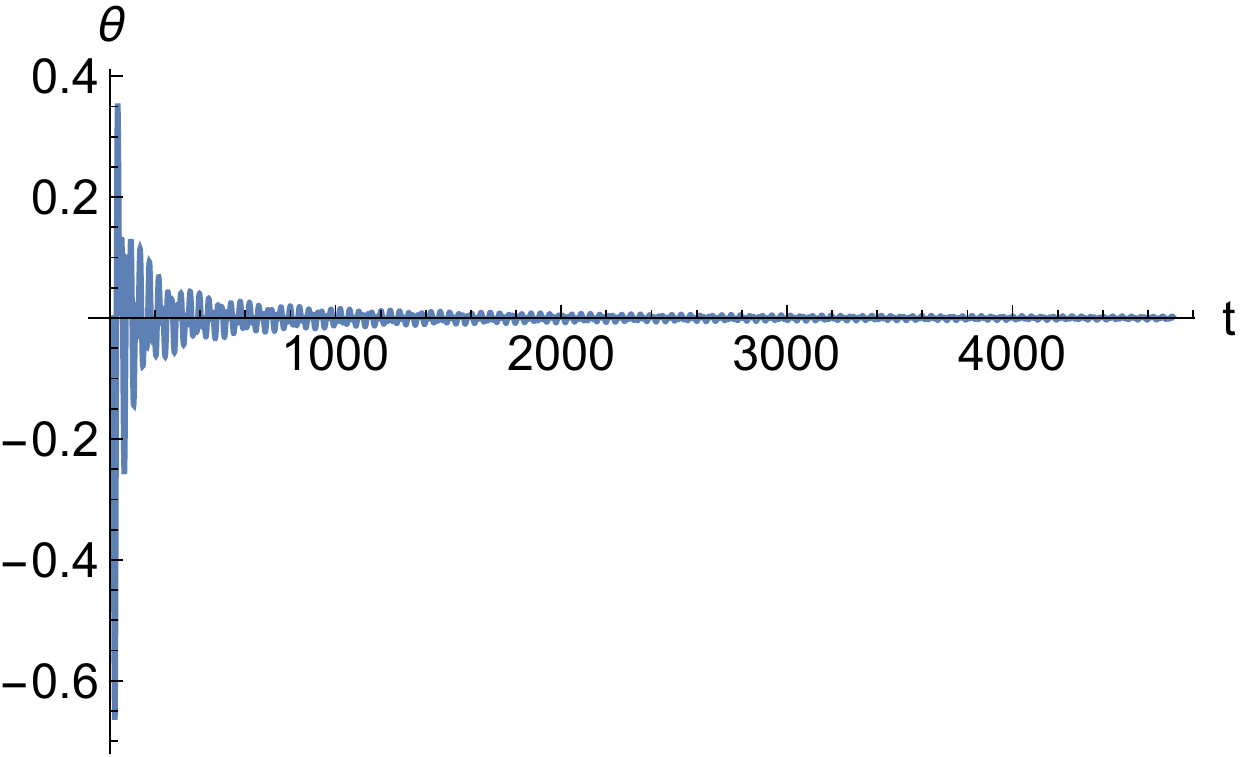} 
 \includegraphics[width=.495\textwidth,clip]{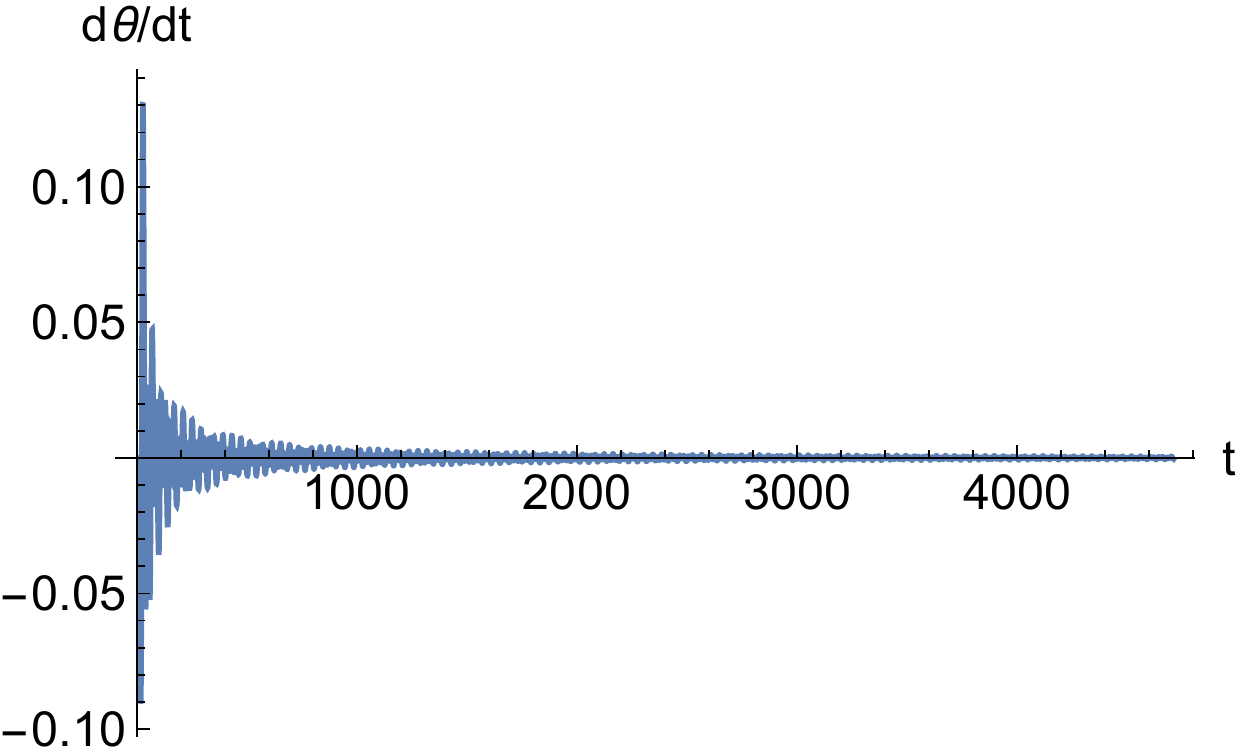} 
 \caption{The phase of the scalar baryon (left) and 
the phase velocity of the scalar baryon (right) with $(\lambda, 1/\Lambda) = (10^{8}, 1/\MP)$, are depicted.
$\dot{\theta}$ and $t$ are measured by using units of $\Treh$ and $1/\Treh$ respectively.} 
\label{fig:potential_large}
\end{figure}

\begin{figure}[ht]
\center
 \includegraphics[width=.495\textwidth,clip]{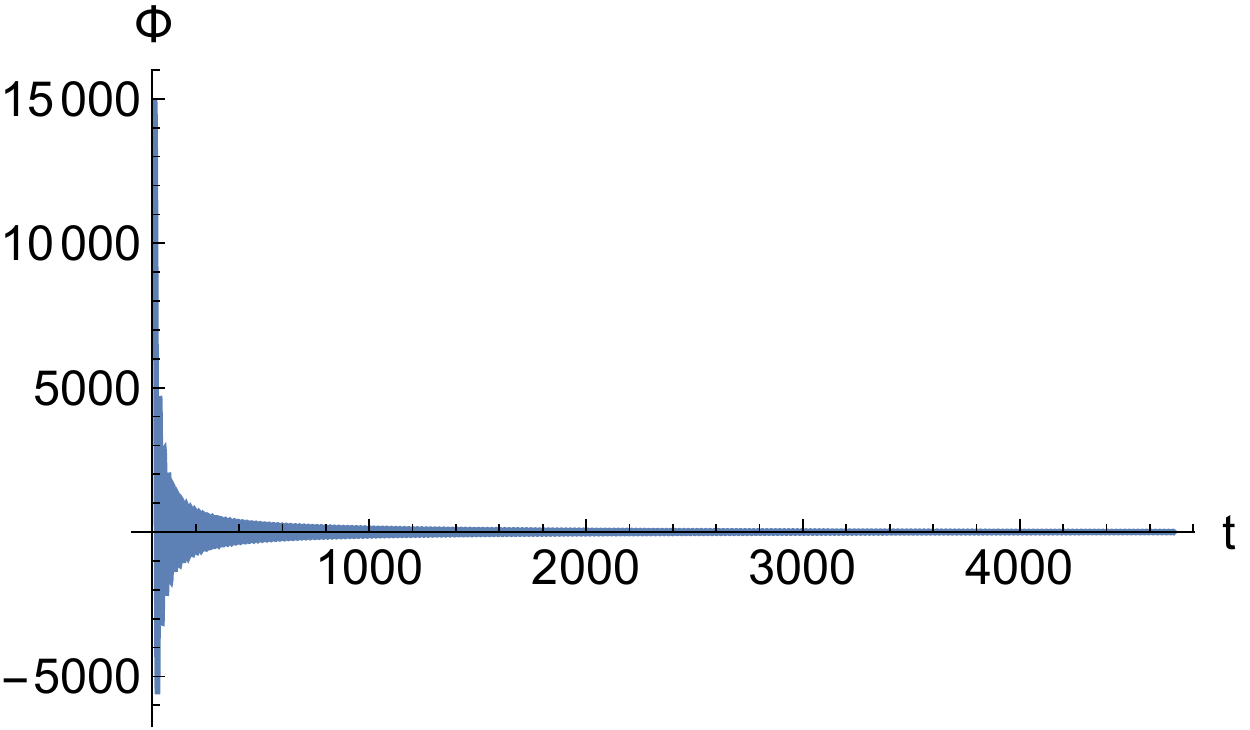}\\ 
 \includegraphics[width=.495\textwidth,clip]{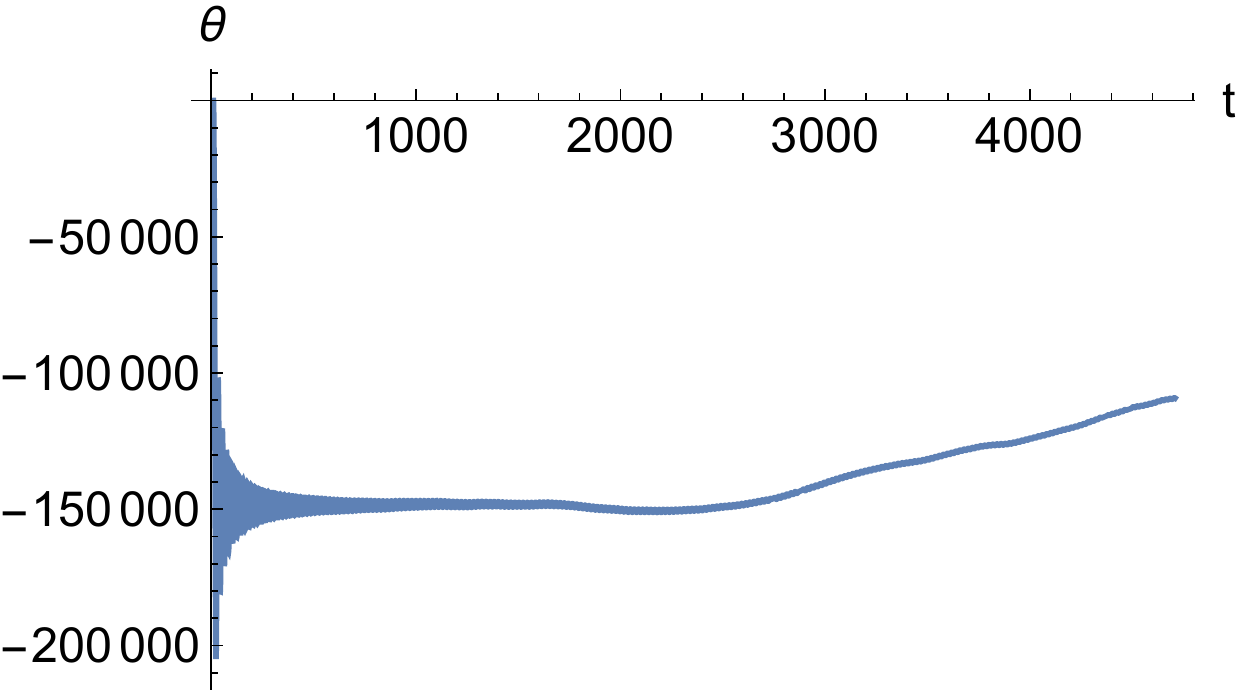} 
 \includegraphics[width=.495\textwidth,clip]{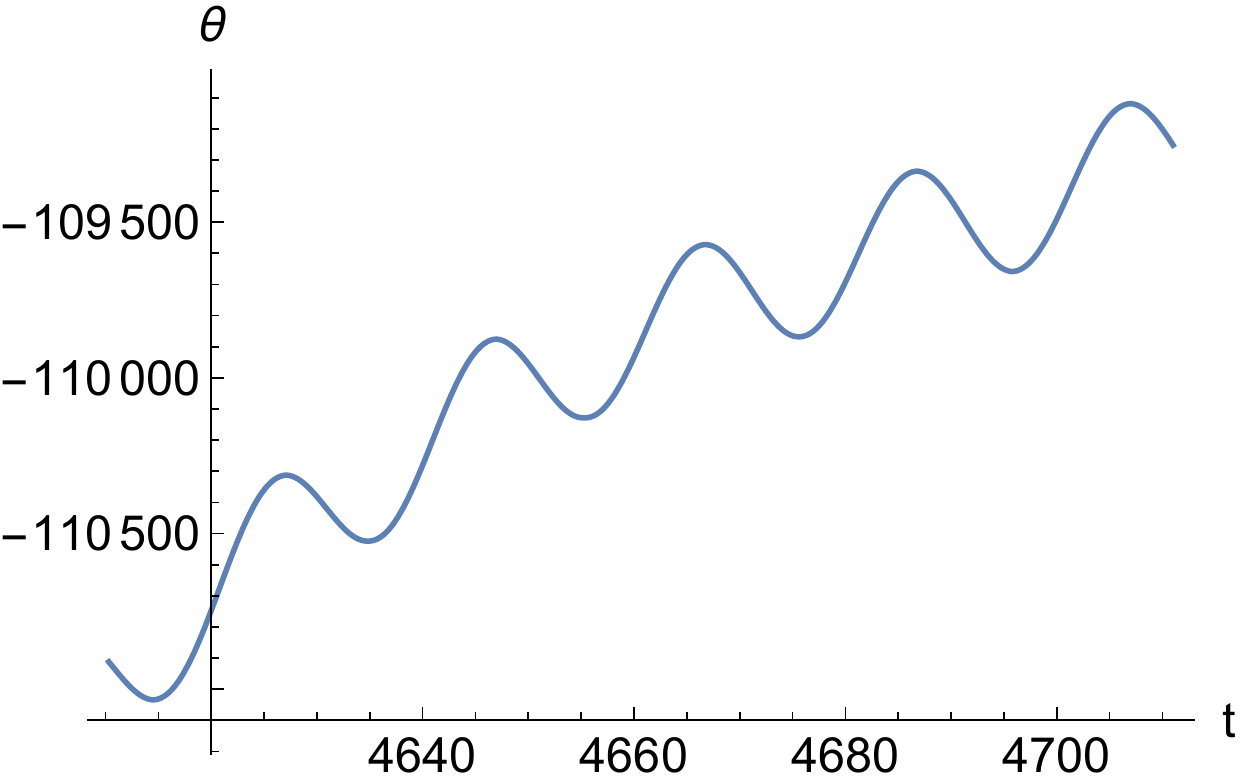} 
 \caption{The time evolution of the inflaton (top), phase of the scalar baryon (bottom left)
 and phase of the scalar baryon at the end of the reheating (bottom right)
with $(\lambda, 1/\Lambda) = (10^{11}, 10/\Treh)$ are plotted.
$\Phi$ and $t$ are measured by using the units $\Treh$ and $1/\Treh$ respectively.} 
\label{fig:result}
\end{figure}

\begin{figure}[ht]
\center
 \includegraphics[width=.495\textwidth,clip]{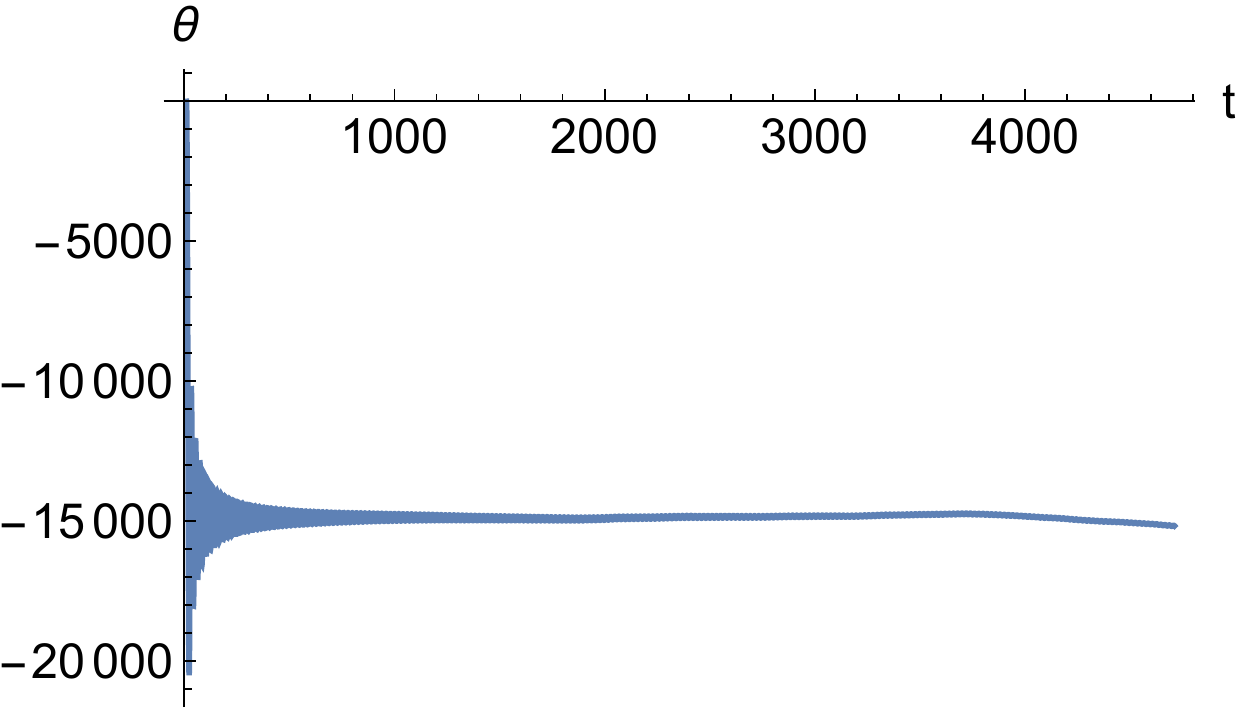}
 \includegraphics[width=.495\textwidth,clip]{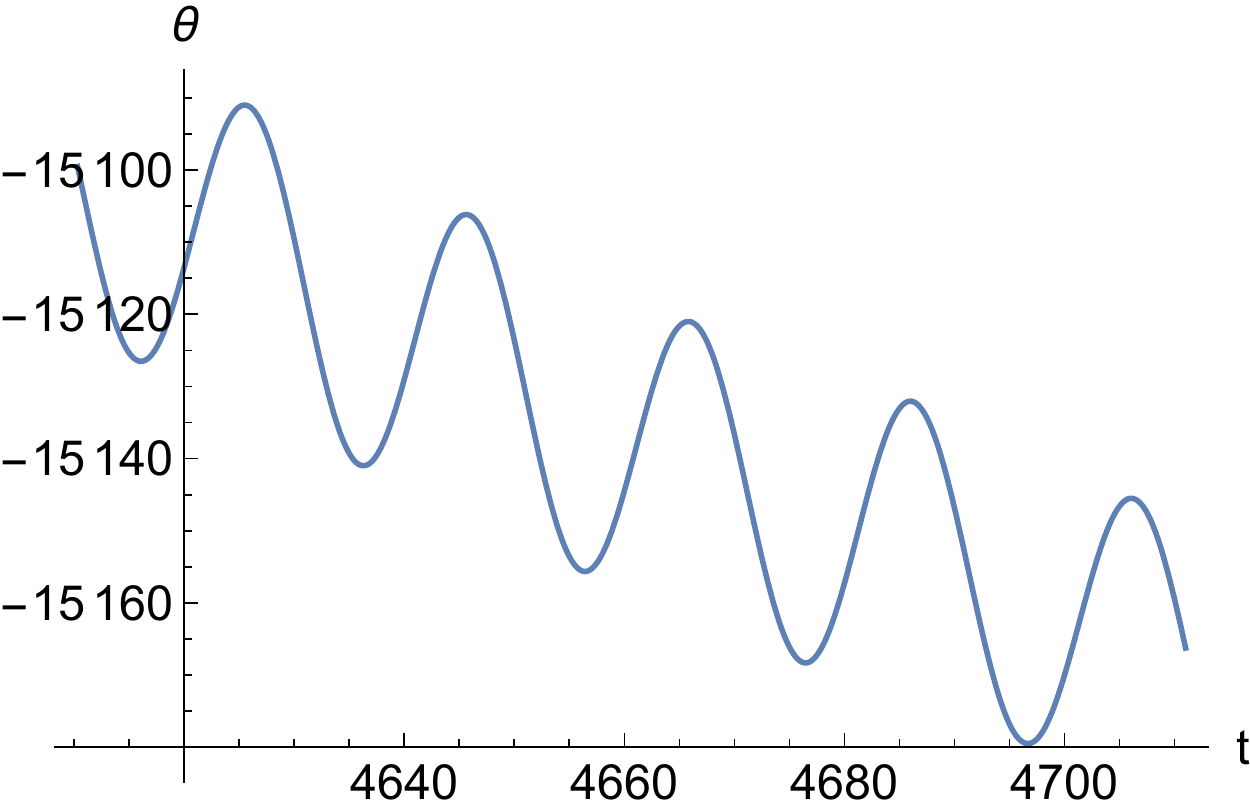}
 \caption{The phase of the scalar baryon throughout reheating (left) and at the end of the reheating (right)
 with $(\lambda, 1/\Lambda) = (10^9, 1/\Treh)$ are illustrated.
 $t$ is measured by using in units of $1/\Treh$.}
\label{fig:monotonous}
\end{figure}

Numerically, we have various such solutions.  Typical examples are depicted in Fig.~\ref{fig:potential_large} - Fig.~\ref{fig:monotonous}.

In Fig.~\ref{fig:potential_large}, no monotonously increasing (directional) motion of $\theta$ appears.  The parameters chosen are $(\lambda, 1/\Lambda)=(10^8, 1/\MP=3 \times 10^{18} \mbox{GeV})$.
On the other hand, monotonously increasing motion appears in Fig.~\ref{fig:result} and Fig.~\ref{fig:monotonous}.  
The parameters in Fig.~\ref{fig:result} are $(\lambda, 1/\Lambda)=(10^{11}, 10/\Treh)$, while in Fig.~\ref{fig:monotonous} we have chosen $(\lambda, 1/\Lambda)=(10^{9}, 1/\Treh)$.  
The sweet spot parameters in Fig.~\ref{fig:result} and Fig.~\ref{fig:monotonous}, giving baryogenesis, shows $\lambda \approx 10^{9\sim 10} \frac{\Treh}{\Lambda}$ satisfies the adjustment condition between the external pushing/driving force and the gravitational force.

The other parameters necessary to carry out the simulation are as follows:
we choose the reheating temperature as $\Treh=10^{14}$ GeV,
the Hubble parameters at the initial and the final stage of the reheating are derived to be $H_i=6.2 \times 10^{12}$ GeV and $H_f=1.4 \times 10^{10}$ GeV respectively, $\phi_r=10^9$ GeV and $\Gamma=1.8 \times 10^{10}$ GeV.
We can then define $t_i=2/(3H_i) $ and $ t_f=2/(3H_f)$ because during reheating the oscillation of the massive inflaton gives rise to an approximately matter dominated epoch.
We see that $t_f - t_i \sim 1/\Gamma$~(a lifetime of the inflaton) holds.
In the simulation, $\theta=\dot{\theta}=0$ is chosen at the beginning of inflation where $\Phi=0.62\, \MP$ and $\dot{\Phi}=0$.


We take time and spaces which are compatible with the end of reheating epoch
$\tilde{t}=(cT_{\mathrm{reh}}) t, (\tilde{x},\tilde{y},\tilde{z})=cT_{\mathrm{reh}}(x,y,z)$,
where $c $ is some numerical constant.
This means that the time in our problem should be measured in the unit of $1/T_{\mathrm{reh}}$, see Fig.~\ref{fig:result} and Fig.~\ref{fig:monotonous}.
 
In terms of this time $\tilde{t}$, the action Eq. (1) is written as follows:
\begin{eqnarray}
\label{eq:rescale}
S=\frac{\phi_r^2}{(c T_{\mathrm{reh}})^2}\int d^4\tilde{x} ~a(\tilde{t})^3 
\times \left[ \frac{1}{2}\theta^{'2}-\left(\frac{\lambda \phi_r^2}{(c T_{\mathrm{reh}})^2}\right) \sin^2\theta+ \frac{1}{2}\tilde{\Phi}^{'2}-\frac{1}{2} \left(\frac{\mu}{c T_{\mathrm{reh}}}\right)^2 \tilde{\Phi}^{2}- \frac{1}{\tilde{\Lambda}}\theta' \tilde{\Phi}' \right],
\end{eqnarray}
where time derivative is written by the prime.  
We understand that a parameter
\begin{eqnarray}
\tilde{\lambda} = \frac{\lambda \phi_r^2}{(cT_{\mathrm{reh}})^2}
\end{eqnarray}
naturally arises in the above Eq.~\eqref{eq:rescale}, and

\begin{eqnarray}
\omega=\frac{\mu}{c T_{\mathrm{reh}}}, ~\tilde{\Phi}=\frac{\Phi}{\phi_r}, ~\mathrm{and}~ \frac{1}{\tilde{\Lambda}}=\frac{\phi_r}{\Lambda}.
\end{eqnarray}

For our choice of parameters in our numerical calculations, $c=1$, $\tilde{\lambda}=(0.01-10)$ (potential of the rotator), $\omega=0.3$ (Inflaton's push), and $\frac{1}{\tilde{\Lambda}}=10^{-10}-10^{-4}$ (CP violation). 
It is difficult to determine $c$, but if we can take $c=10$, $\tilde{\lambda}=(0.0001-0.1)$. 

In the following when we rewrite the model in terms of the forced pendulum.   
A concern is that the scalar potential takes a sinusoidal form, having maxima as well as minima.  At the maxima the fluctuations of the $\phi$ field behaves like a tachyon.  This is an important issue to clarify in our solution from the point of view of  tachyonic reheating~\cite{Greene:1997ge} and reheating by parametric resonance~\cite{Kofman:1997yn}, which will be discussed below. 


Such a monotonously increasing (directional) solution is known as a ``phase locked state" in the forced pendulum scenario.  
Pedersen {\it et al.} clarify such solutions in the study of the chaotic behavior of the electric current passing through Josephson junctions~\cite{Pedersen:1980}.  Since the notation adopted by D'Humieres {\it et al.} in Ref.~\cite{D'Humieres:1982} is convenient, we change the variables and parameters accordingly.\footnote{$\Theta=2\theta, \tau =\sqrt{2p}\left[(t-t_i)-\frac{\pi}{\mu}\right],~\omega=\frac{\mu}{\sqrt{2p}}\,,~ Q =\sqrt{2p}\frac{1}{f}=\sqrt{\frac{p}{2}}\,t_f\,,~\mbox{and} ~\gamma =\frac{q(t_f)}{p}\,$.}  Then, the equation of motion for $\theta$ becomes 
\begin{equation}
\ddot{\Theta} +  \frac{1}{Q}\dot{\Theta} + \sin\Theta \;=\; \gamma\cos(\omega \tau)\;, 
\label{EOM}
\end{equation}
where the derivative with respect to $\tau$ is still expressed by a dot. 
The phase locked state is a solution of the form
\begin{eqnarray}
\Theta(\tau) \;=\; \Theta_n + n (\omega \tau-\phi_0) - \alpha \sin (\omega \tau -\phi_0)\;.
\label{eq:phase_locked_state}
\end{eqnarray}
The condition for this ansatz to fulfil the equation of motion as well as the stability condition against external perturbations are known~\cite{Pedersen:1980, D'Humieres:1982}.  
For example $\alpha$ is the zero point of the $n$-th Bessel function.  However, $n$ is not necessary to be an integer.
This is a solution in which the frequency in the driving force $\omega$ coincides with the frequency of modulation in $\alpha \sin (\omega \tau -\phi_0)$ for the directed motion of $\Theta$, but more general solutions can be obtained.

The numerical solution follows this phase locked state with the modulation frequency equal to the frequency of the inflaton the period of which is $2\pi/\mu \sim 20$ in our units.
From Fig.~\ref{fig:result} and Fig.~\ref{fig:monotonous}, we can read off $n=110$ and $-4.2$, respectively.

The amplitude of the phase locked state increases linearly in time, while the amplitude driven by the parametric resonance increases exponentially in time.  The difference comes from the fact that in the former case, the external force is modulated in time from outside, while in the latter case the frequency of the oscillator is modulated.  The latter case of the parametric resonance is used to realize the explosively rapid preheating by Kofman, Linde and Starobinsky~\cite{Kofman:1997yn}.  In our case of the Cohen-Kaplan type interaction the phase locked state appears. 


Here we will comment on the tachyonic reheating in addition to the parametric resonance.  Both reheating mechanisms can be understood through the instability of the classical solution of scalars using the Mathieu differential equation \cite{Greene:1997ge}.  As was stated above the instability of our phase locked state is also examined using the Mathieu equation.

We introduce the fluctuation of $\Theta$ by  $\Theta_*+\delta\Theta$, where $\Theta(t)_*$ is the solution of the phase locked state given in Eq.~(\ref{eq:phase_locked_state}).  Then, the fluctuation satisfies the Mathieu differential equation
\begin{eqnarray}
\ddot{\delta \Theta} + \cos(\Theta(t)_*) \delta{\Theta}=0,
\end{eqnarray}
in the linear approximation.  The $\cos(\Theta_*)$ term in the above is a remnant of the sinusoidal potential. The tachyonic problem (the fluctuation $\delta \Theta$ behaves tachyonic during some period) is taken into account as the instability of the solution, that is, the exponential increase of the fluctuation $\delta \Theta$.  Reheating achieved by parametric resonance and that by production of tachyons use the parameter regions in which the ``instability'' of the Mathieu differential equation takes place.  The instability gives the exponential growth of the field.  On the other hand, in our phase locked state, the field increases linearly in time and the Mathieu differential equation appearing not in the equation of motion itself, but in the equation determining the instability.  We, of course, choose the stability regions of the Mathieu differential equation.  Therefore, our solution stands at the fringe of chaotic motion.  However, the solution relevant to baryogenesis is that (the phase locked state) increasing linearly in time. 


As for the estimation of $\eta_B$, if the entropy produced during the reheating is $s_\mathrm{reh}=(2\pi^2/45) g^* T^3_\mathrm{reh}$ with the SM value of $g^*=106.5$, we have $\eta_B^\mathrm{reh} \approx 0.01n \times \left(\mu \phi_r^2/T^3_\mathrm{reh}\right)$.  After the reheating ends, EW baryogenesis may dilute the value to $\eta_B=\frac{28}{79} \eta_B^\mathrm{reh}$, and hence,
in order to reproduce the observed value of $\eta_B$, $n \sim 770$ with $\phi_r=10^9$ GeV and $\Treh=10^{14}$ GeV.

This ratchet baryogenesis is analogous to the forced pendulum.  It is well known that the behavior of the latter is sometimes chaotic.  So, the sweet spot solution of the phase locked state which can give the baryogenesis may be on the fringe of chaos.  If we remember, however, that the mechanism is also analogous to the ratchet model of molecular motors, there may exist stable regions of the parameter space, 
since we can expect living organisms to have evolved to possess stable mechanisms for directional motion.  
Unfortunately, if there is not a large parameter region to produce the stable phase locked states, we have to average statistically over the various solutions with different initial conditions.  A way of doing this averaging over the solutions is to use a Poincar\'{e} map.  A Poincar\'{e} map is the the discretized collection of orbit points $(x(\tau), y(\tau))$ at $\tau=\frac{2\pi}{\omega}\times (\mbox{integers})$, where various initial conditions for $x=\theta$, and $y=\dot{\theta}$ are taken.  So, the distribution of the points on the map can be understood as a ``coarse grained distribution function" $f_{dis}(x, y)$ of the dynamical system in the phase space $(x, y)$.

\begin{figure}[ht]
\center
\includegraphics[width=.495\textwidth,clip]{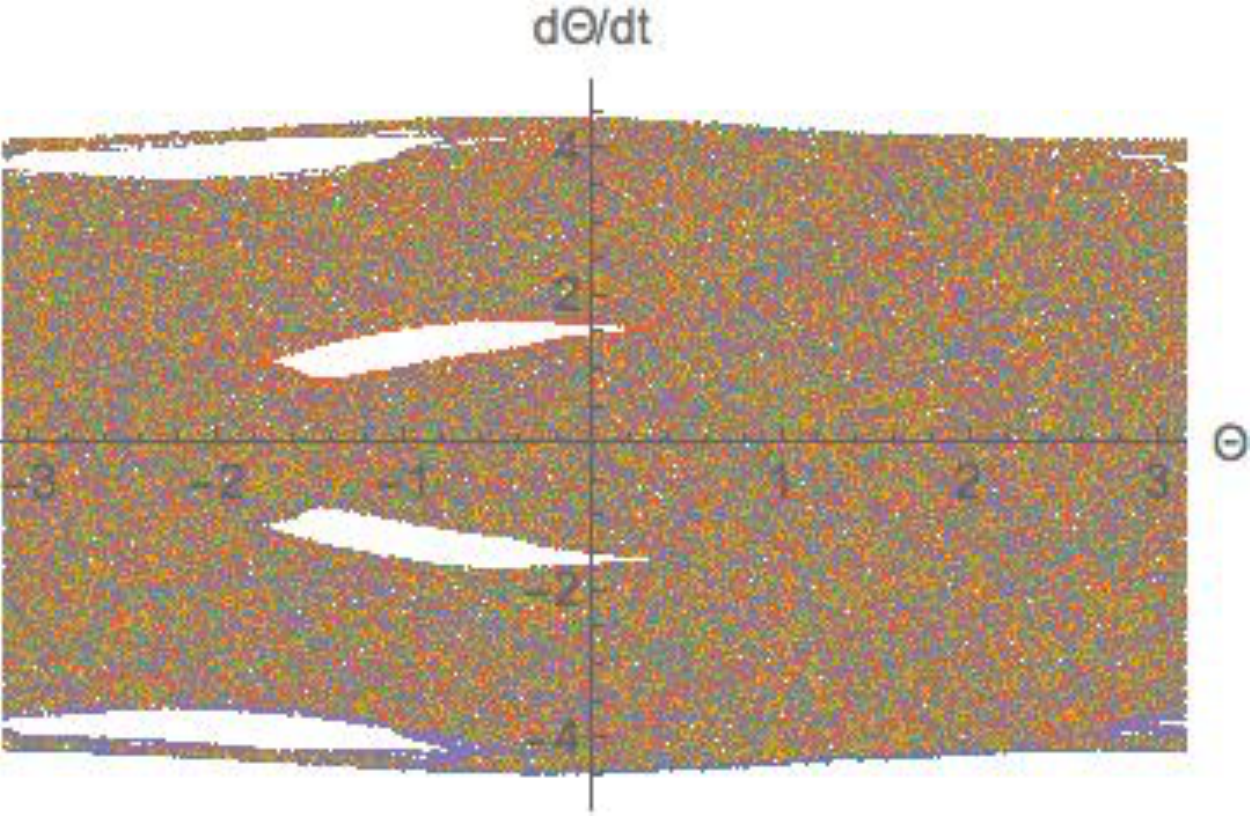}
 \caption{A Poincar\'{e} map is depicted without friction with $(\gamma, \omega) = (2, 1)$. 
 20 initial values $(x_j(0), y_j(0))$ $(j=0,1,\cdots 19)$ are taken to be $(-\pi/2 + j\pi/20, 0)$.}
\label{fig:poincare_wo_fric}
\end{figure}

\begin{figure}[ht]
\center
\includegraphics[width=.495\textwidth,clip]{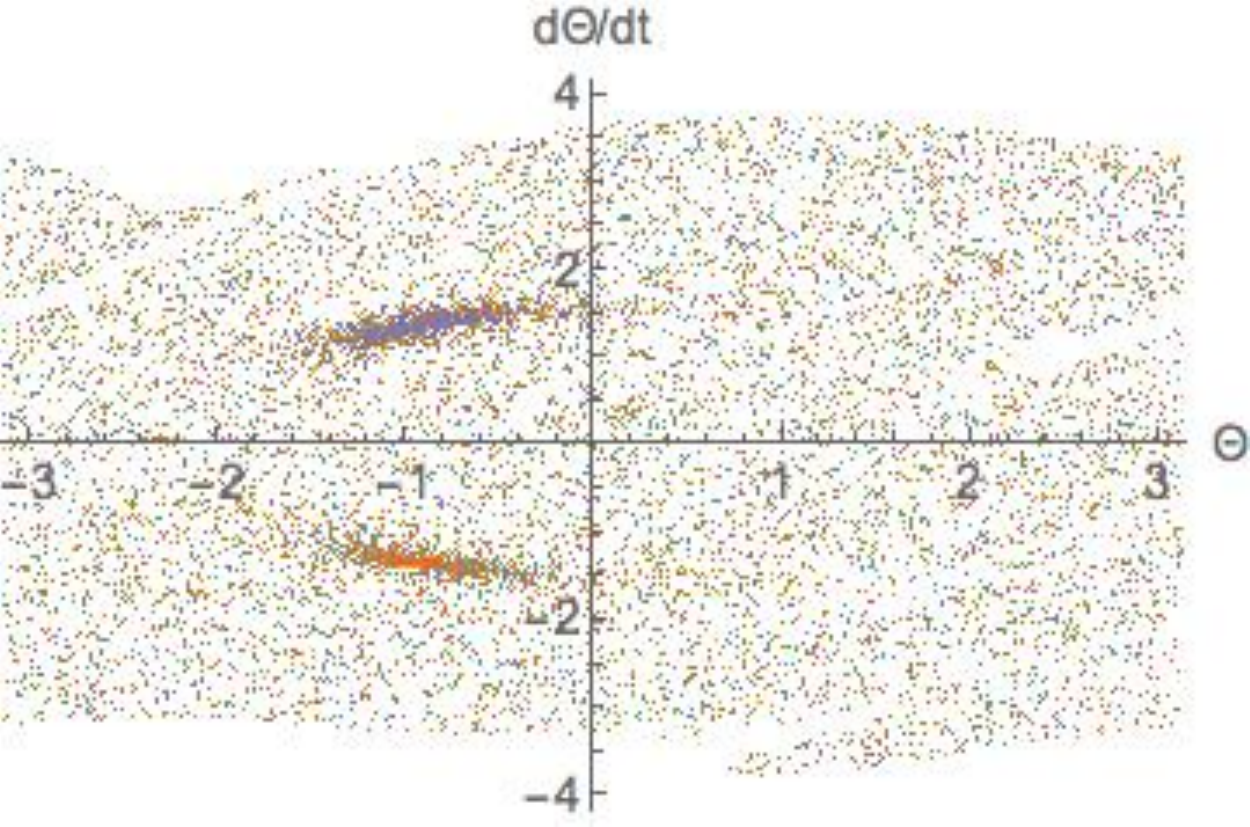}
 \caption{A Poincar\'{e} map is depicted with friction of $f_\mathrm{fric}=1/Q=0.005$ with $(\gamma, \omega) = (2, 1)$.
  20 initial values $(x_j(0), y_j(0))$ $(j=0,1,\cdots 19)$ are taken to be $(-\pi/2 + j\pi/20, 0)$.}
\label{fig:poincare_w_fric}
\end{figure}

Now the averaged baryon number reads
\begin{equation}
\langle n_B \rangle = \phi_r^2 \langle y \rangle \;=\; \phi_r^2\;\frac{\displaystyle \sum_{j~(\mathrm{initial~conditions})} \int dx_j\,dy_j\,y_j f_{\mathrm{dis}}(x_j, y_j) }{\displaystyle \sum_{j~(\mathrm{initial~conditions})} \int dx_j\,dy_j\,f_{\mathrm{dis}}(x_j, y_j) }
\end{equation}
It is interesting to compare the following two Poincar\'{e} maps: 
Fig.~\ref{fig:poincare_wo_fric} depicts evolution without the friction $f_\mathrm{fric}=1/Q=0$, while 
Fig.~\ref{fig:poincare_w_fric} depicts evolution with the friction term of $f_\mathrm{fric}=1/Q=0.005$.  
Here we take the parameters as $\gamma=2$, $\omega=1$ in Eq.~\eqref{EOM} as a demonstration.
20 initial values $(x_j(0), y_j(0))$ $(j=0,1,\cdots 19)$ are taken to be $(-\pi/2 + j\pi/20, 0)=$
$(-\pi/2, 0)$, $(-\pi/2 + \pi/20, 0)$, $\cdots$, $(-\pi/2 + 19\pi/20, 0)$.


The averaged $\langle n_B \rangle$ over the initial conditions is $-0.095$ $\approx 0$ for Fig.~\ref{fig:poincare_wo_fric}, but is $0.31$ for Fig.~\ref{fig:poincare_w_fric}. 
From this, we will learn that the friction term $f_\mathrm{fric}$ introduced for producing a non-thermal-equilibrium state may be essential
for baryogenesis to survive after averaging over the various initial conditions, when considering the generation of baryon number near the end of reheating.  
Possibly, the averaged value $\langle n_B \rangle$ may peak strongly at a certain choice of parameters, since we expect biological motors to rotate like a classical motor with little statistical fluctuations.


So far we have identified the baryon number density $n_B$ to that of the scalar baryon $n_{B}(\mathrm{scalar})=\phi_r^2 \dot{\theta}$.  
However, the actual baryon number density in our universe is that of the fermionic matter made up of $u$ and $d$ quarks.  
Therefore, the baryon number density of the scalar baryon must be converted to that of fermionic baryons.  

There are multiple ways to accomplish this conversion such as introducing the Yukawa interaction 
\begin{equation}
\Delta\mathcal{L}_{\mathrm{int}} 
\;=\; \phi\,\overline{Q}L + \mathrm{h.c.}
\end{equation}
where $Q$ and $L$ are fermions with baryon number 1 and 0, respectively.
When the scalar baryons decay into $Q\overline{L}$ pairs, the baryon number in the scalars will be
efficiently converted into that in fermions.
Note that this interaction, used by Dolgov and Freese in Ref.~\cite{Dolgov:1994zq},
is only a toy example since the $Q$ field in this expression cannot carry $SU(3)$ color,
but is sufficient for illustrative purposes.

If we insist on using SM fields directly for the scalar-fermion conversion, 
possible $SU(3)_C\times SU(2)_L\times U(1)_Y\times B$ invariant
interactions would be the dimension 7 operators that can be written schematically as
\begin{eqnarray}
\Delta\mathcal{L}_{\mathrm{int}} 
& = & \dfrac{1}{\Lambda^3}\,\phi^* \Bigl[ (u_L d_L d_L) \nu_L - (u_L u_L d_L) e_L \Bigr] + \mathrm{h.c.}
\cr 
& & \mbox{or}\quad
\dfrac{1}{\Lambda^3}\,\phi^* (u_R d_R d_R \nu_R) + \mathrm{h.c.}
\cr 
& & \mbox{or}\quad
\dfrac{1}{\Lambda^3}\,\phi^* (u_R u_R d_R e_R) + \mathrm{h.c.}
\end{eqnarray}
where color and spinor indices have been suppressed.
If squarks are present in the theory, we could also write dimension 5 interactions of the form
\begin{equation}
\Delta\mathcal{L}_{\mathrm{int}} 
\;=\; 
\dfrac{1}{\Lambda}\,\phi^* (\tilde{u}_R d_R d_R) + \mathrm{h.c.}
\qquad\mbox{or}\qquad
\dfrac{1}{\Lambda}\,\phi^* (u_R d_R \tilde{d}_R) + \mathrm{h.c.}
\end{equation}
The simplest and most realistic model, however, may be to reinterpret our baryon number as lepton number,
and introduce the lepton-number preserving dimension 4 interaction
%
\begin{equation}
\Delta\mathcal{L}_{\mathrm{int}} 
\;=\; 
\phi^* \overline{\nu_R^c}\nu_R^{\phantom{c}} + y H^* \bar{\nu}_R \nu_L+ \mathrm{h.c.}\;,
\end{equation}
%
where $H$ is the standard model Higgs, and $y$ is a Yukawa coupling, that is, the decay of the scalar baryon/lepton would produce a $\nu_R\nu_R$ pair, and $\nu_R$ decays into $\nu_L+ H^*$.
This same interaction can be used to generate a large Majorana mass for $\nu_R$ when $\phi$ obtains the VEV 
$\langle\phi\rangle = \phi_r/\sqrt{2}$, from which we can obtain small Majorana masses for
the (mostly) left-handed active neutrinos via the seesaw mechanism \cite{Yanagida:1979as,Yanagida:1980xy,Ramond:1979py}.
The lepton number generated in this way can then be converted to baryon number later via the $B-L$ conserving sphaleron process \cite{Kuzmin:1985mm,Trodden:1998ym}
as in leptogenesis scenarios \cite{Fukugita:1986hr}.
See also Ref.~\cite{Sugamoto:1982cn}.
More explicitly, the following conversion occurs, $\nu_R \to \nu_L + H^*,  \nu_L \to \overline{(u_Lu_Ld_L)}$, for example.

Alternatively, one could introduce the following dimension 6 interaction between the scalar and fermionic baryon number currents:
\begin{equation}
\Delta\mathcal{L}_{\mathrm{int}} \;=\;
\dfrac{i}{\Lambda^2}\bigl(\phi^* \!\stackrel{\leftrightarrow}{\partial}_\mu\!\phi\bigr)~\frac{1}{3}\bigl(\bar{u}\gamma^\mu u +\bar{d}\gamma^\mu d\bigr) \;,
\label{interacton of scalar baryon and matter}
\end{equation}
where $u$ and $d$ are four component Dirac fields.  
Identifying the charges carried by the scalar and fermionic currents requires the existence of a term in the interaction Lagrangian
which would require both $\phi$ and the quark fields to be transformed at the same time, which we will not show explicitly.
Rewriting $\phi = \dfrac{1}{\sqrt{2}}\phi_r e^{i\theta}$
and ignoring the spatial dependence of $\theta$ we obtain
\begin{equation}
\Delta\mathcal{L}_{\mathrm{int}} \;=\;
-\dfrac{\phi_r^2}{\Lambda^2}\dot{\theta}~\frac{1}{3}\bigl(u^{\dagger}u + d^{\dagger} d\bigr)\;.
\end{equation}
This is also the interaction term used in the baryogenesis model of Cohen and Kaplan \cite{Cohen:1987vi}.
Since the baryon number density of the ordinary matter is $n_B(\mathrm{matter})=\frac{1}{3} \left(u^{\dagger}u+d^{\dagger}d \right)$, its coefficient  can be considered to be the chemical potential $\mu_B$ for ordinary baryon number, {\it i.e.} $\mu_B=-\dfrac{\phi_r^2}{\Lambda^2}\dot{\theta}$.  
This chemical potential gives opposite ``energy" to the matter and antimatter.  Therefore, in the equilibrium state 
at temperature $T_{\mathrm{reh}}$, the baryon number density of ordinary matter is given by
\begin{equation}
\langle n_B \rangle_{\mathrm{matter}}
\;=\; -\frac{2}{9}~ \phi_r^2 \dot{\theta} \left(\frac{T_{\mathrm{reh}}}{\Lambda}\right)^2
\;=\; -\frac{2}{9} n_B(\mathrm{scalar}) \left(\frac{T_{\mathrm{reh}}}{\Lambda}\right)^2.
\end{equation}
In our parameter choice $\Lambda=(0.1\sim 1) T_{\mathrm{reh}}$, and hence the produced baryon number of ordinary matter is about the same as that of the scalar baryon.

In this paper, we have made the simplifying assumption of a uniform isotropic universe and have ignored the spatial dependence of the scalar fields $\Phi$ and $\phi$, and consequently that of $\theta$, in our analysis.
In reality, as the universe expands, different parts of the universe will lose causal contact with each other, and naturally lead to the evolutions of $\Phi$ and $\theta$ picking up
spatial dependences.
In particular, if $\theta$ settles into different vacua with different winding numbers in different domains, domain walls will form at the boundaries creating bubbles of different vacua.
Such domain walls should not be stable since that would contradict observation \cite{Cohen:1997ac} so the bubbles must annihilate themselves sufficiently quickly.
However, when the domain walls move, $\theta$ will necessarily `unwind' and lead to 
non-zero $\dot{\theta}$, generating either baryon or anti-baryon number.
This, and other effects, will be included in a more detailed and complete analysis of our model
to be presented in a forthcoming paper.

\vspace{0.5cm}
A preliminary stage of this work was presented at the SUSY 2015 conference held in August 2015 \cite{Yamashita:2015}.


\section*{Acknowledgements}
K. Y. acknowledges support for a long-term stay in Virginia 
from the Program for Leading Graduate Schools of Ochanomizu University; 
she thanks Theoretical Particle Physics \& String Theory group at Virginia Tech 
for the warm hospitality while this work was conducted.

%
%
\bibliographystyle{ptephy}
\bibliography{arXiv_v3_baryogenesis}
\end{document}